\documentclass[reqno]{amsart}
\usepackage{epsf}
\def\be{\begin{equation}}
\def\ee{\end{equation}}
\def\ba{\begin{array}}
\def\ea{\end{array}}
\def\bea{\begin{eqnarray}}
\def\eea{\end{eqnarray}}
\def\drm{{\mathrm d}}
\def\erm{{\mathrm e}}
\def\dps{\displaystyle}

\def\ov{\overline}

\def\gr{\mbox{\boldmath $\nabla$} }
\def\Abvec{\mbox{\boldmath $A$}}
\def\Ebvec{\mbox{\boldmath $E$}}
\def\Hbvec{\mbox{\boldmath $H$}}
\def\l{\lambda}
\def\fbvec{\mbox{\boldmath $f$}}
\def\sbvec{\mbox{\boldmath $s$}}
\def\kbvec{\mbox{\boldmath $k$}}
\def\xbvec{\mbox{\boldmath $x$}}

\def\la{\gr^2}
\def\di{\gr \, {\cdot} \,}
\def\ro{\gr \, {\times} \,}

\def\cibvec{\mbox{\boldmath $I$}}

\begin{document}

\vspace{-4truecm} %
{}\hfill{DSF$-$29/2007}%

%{}\hfill{physics/yymmnnn}%
\vspace{1truecm}

\title[An unknown story: Majorana and the Pauli-Weisskopf
theory]{An unknown story: Majorana and the Pauli-Weisskopf scalar
electrodynamics}

\author{S. Esposito}
\address{{\it S. Esposito}: Dipartimento di Scienze Fisiche,
Universit\`a di Napoli ``Federico II'' \& I.N.F.N. Sezione di
Napoli, Complesso Universitario di M. S. Angelo, Via Cinthia,
80126 Napoli ({\rm Salvatore.Esposito@na.infn.it})}%

\begin{abstract}
An account is given of an interesting but unknown theory by
Majorana regarding scalar quantum electrodynamics, elaborated
several years before the known Pauli-Weisskopf theory. Theoretical
calculations and their interpretation are given in detail,
together with a general historical discussion of the main steps
towards the building of a quantum field theory for
electrodynamics. A possible peculiar application to nuclear
constitution, as conceived around 1930, considered by Majorana is
as well discussed.
\end{abstract}

\maketitle

%keywords: Ettore Majorana; Pauli-Weisskopf theory; Quantum
%Electrodynamics; Quantum Field Theory; Nuclear constitution

%----------------------------------------------------------------

\section{The known story of early Quantum Electrodynamics}

\begin{quote}
Until a few years ago it had been impossible to construct a theory
of radiation which could account satisfactorily both for
interference phenomena and the phenomena of emission and
absorption of light by matter. The first set of phenomena was
interpreted by the wave theory, and the second set by the theory
of light quanta. It was not until in 1927 that Dirac succeeded in
constructing a quantum theory of radiation which could explain in
an unified way both types of phenomena \cite{FermiRMP}.
\end{quote}

\subsection{First steps in Quantum Field Theory}

\noindent With these words, according to what was later published
\cite{FermiRMP}, Enrico Fermi opened the first of his renowned
lectures on the ``Quantum theory of radiation'', delivered at the
University of Michigan - Ann Arbor in 1930. The basic problem for
the building up of a quantum electrodynamics was, in fact, the
well recognized dichotomy concerning the description of particles
(electrons) in a wave-mechanical way, as well as the description
of light quanta (photons) in the framework of the Maxwell theory
of electromagnetism.

The first preliminary step towards the clarification of such an
issue, just after the appearance of Quantum Mechanics, was carried
out by Max Born, Werner Heisenberg and Pascual Jordan in 1926
\cite{Jordan1926}. By studying the problem of the vibrations of a
one-dimensional string fixed at both sides, Jordan interpreted the
quantum numbers of individual oscillators as the number of light
quanta with given frequency in each cell (in the Bose-Einstein
sense). This one-to-one correspondence proved to give the correct
fluctuation law for cavity radiation,  thus starting to shed some
light in the puzzle mentioned earlier. The method developed by
Jordan was then applied to a general theory of electromagnetic
radiation by Paul A.M. Dirac in the fundamental papers
\cite{Dirac1927a,Dirac1927b} quoted above by Fermi.
\begin{quote}
The key viewpoint in the Dirac theory of radiation is to consider
a light emitting or absorbing atom and the radiation in the space
[around it] as a single system. This is described, apart from the
atomic coordinates, in terms also of some other variables which
specify the values of the potentials of the electromagnetic field
in any point of the space. If, in order to simplify the
discussion, we think at the radiation inside a cavity, we know
that the radiation [field] may vibrate according to certain
characteristic frequencies, depending on the form and dimension of
the cavity. In this case, for each fundamental vibration, one can
take the value of the electric field at a given time in one of the
antinodes (or a quantity proportional to it) as the coordinate of
the radiation field, so that we have one coordinate for each
fundamental frequency. If we allow the walls of the cavity to tend
to the infinity, the fundamental vibrations of the radiation tend
to given asymptotic distributions in such a way that, in the
limiting case, any influence of the walls of the cavity on the
phenomenon ceases, and we are led to the case of the free space.
\cite{FNM52}
\end{quote}

\noindent Dirac described the system at hand in hamiltonian terms,
and the wavefunction of the perturbed hamiltonian system,
satisfying the time-dependent Schr\"odinger equation, was expanded
in terms of the unperturbed wavefunctions (plane waves). Here the
main point is that the expansion coefficients $a_r,a_s^\ast$ (or
some other coefficients $b_r,b_s^\ast$ related to $a_r,a_s^\ast$
by a given transformation), taken as canonical conjugate
variables, are
assumed to be q-numbers satisfying the commutation relations %
\be %
[ a_r , a_s^\ast ] = \delta_{r,s}, \qquad \qquad [ a_r , a_s ] = [
a_r^\ast , a_s^\ast ] = 0 . \label{e1}
\ee %
This further or ``second'' quantization of such operators proved
useful in demonstrating that the Hamiltonian considered by Dirac
was for an assembly of particles obeying the Bose-Einstein
statistics. The hamiltonian interaction term between the radiation
field and the atom was shown to have the same form of that
pertaining to an assembly of light quanta interacting with the
atom. ``There is thus a complete formal reconciliation between the
wave and light quantum points of vie'' \cite{Dirac1927b}. The
Dirac theory was soon recognized to describe correctly and
satisfactorily the emission and absorption  of light, the
dispersion phenomena and, more in general, all the phenomena
related to energy exchange between light and matter.

Few months later the appearance of the Dirac papers, Jordan and
Oskar Klein \cite{JordanKlein1927} extended the Dirac theory to
the case of a system of massive bosons and, finally, Jordan and
Eugene P. Wigner in 1928 \cite{JordanWigner1928} obtained a
quantization method for fermions. They showed that, for electrons,
the field cannot be expanded in terms of operators satisfying the
commutation relations in Eq. (\ref{e1}), and introduced
anticommutators for the field variables in order to take into
account the Pauli exclusion principle. The use of canonical
commutation or anticommutation relations for the operators
$a,a^\ast$ raised the problem of Lorentz invariance of the
second-quantized theory. For the charge-free case, the problem of
a relativistically invariant formulation of the Dirac
theory\footnote{In the Dirac theory, the electromagnetic potential
was separated into a radiation field $\Abvec$ and a static Coulomb
potential $\varphi$ in a way that gauge and Lorentz invariance was
not preserved, on the contrary to what manifestly happened in the
classical electrodynamics theory.} was considered by Jordan and
Wolfgang Pauli later in 1928 \cite{JordanPauli1928} (addressing,
in addition, also the problem with the infinite zero-point energy)
who, in a certain sense, generalized the Dirac method by
quantizing directly the field strengths instead of the potentials.
As Dirac, they decomposed the radiation field in its harmonic
components, and quantization conditions similar to those for the
harmonic oscillator were applied to each of these components. The
commutation relations between the electric and magnetic field
components were derived in such a way that relativistic invariance
holds, showing that, {\it inter alia}, the quantized
electromagnetic field propagates at the velocity of light and that
electric and magnetic fields may be measured to arbitrary accuracy
at two points with a space-time separation.

\subsection{Towards a Quantum Field Theory for Electrodynamics}

The formalism introduced by Jordan and Pauli was quite general,
but it applied only to the charge-free case, whilst Dirac
considered, though in an approximate way, also the presence of
electric charges. However, as explicitly stated, the Dirac
expansion in unperturbed plane waves assumed the electromagnetic
field to be only a radiation field, so that the results applied
only far from the charges generating the field. Thus the need for
a further generalization, taking into account, for example, the
action between two neighbouring atoms or two electrons in the same
atom, was felt by many authors. In 1929 Heisenberg and Pauli
\cite{HP1929a,HP1929b}, on one hand, and Fermi
\cite{FNM50,FNM52,FNM64,FermiRMP} on the other hand studied and
solved in different ways this problem. The key point was just to
recognize that ``the Hamiltonian representing the natural quantum
translation of the classical electrodynamics, in the sense given
by the principle of correspondence, is simply obtained by adding
to the Hamiltonian of the Dirac radiation theory a term
representing the electrostatic energy of the electric charges''
\cite{FNM64}. The two approaches were quite different in the
formalism used.
\begin{quote}
A general quantum theory of the electromagnetic field was
constructed by Heisenberg and Pauli by a method in which the
values of the electromagnetic potentials an all the points of
space are considered as variables.
\\
Independently [Fermi] proposed another method of quantization of
the electromagnetic field starting from a Fourier analysis of the
potentials. though Heisenberg and Pauli method  puts in evidence
much more clearly the properties of relativistic invariance and is
in many regards more general, [...] [Fermi method] is more simple
and more analogous to the methods used in the theory of radiation.
\cite{FermiRMP}.
\end{quote}
The method adopted by Heisenberg and Pauli was, in fact, very
general and based on lagrangian field theory. After the writing of
a Lagrangian density $L$ for the fields $\varphi_a$ describing a
given physical system, and the calculation of the canonical
momenta $\Pi_a$ conjugate to the fields $\varphi_a$ and the
Hamiltonian density $H$ of the system, the quantization is
introduced by requiring that $\varphi_a$ and $\Pi_a$ be
q-numbers satisfying the relations%
\be %
[ \Pi_a, \varphi_a ] = \frac{\hbar}{i} , \label{e2}
\ee %
where the commutators (for bosons) or anticommutators (for
fermions) are evaluated at equal times for different spatial
points. This general formalism was applied to the cases of a Dirac
field and of the electromagnetic field, by studying the
fundamental invariance and conservation laws. However, a major
difficulty in applying such a formalism to the electromagnetic
field was that the canonical momentum conjugate to the scalar
potential is zero so that, for this field component, Eq.
(\ref{e2}) cannot apply, the left-hand-side being identically
zero. A first attempt to solve this problem \cite{HP1929a} was to
introduce a Lagrange multiplier in the original Lagrangian
density, preserving gauge and relativistic invariance, this
parameter being set to zero at the end of any calculation.
Subsequently Heisenberg and Pauli demonstrated that, in a
relativistically invariant manner, one can always choose a gauge
where the scalar potential is zero, $\varphi=0$, so that
calculations can be carried out in the Coulomb gauge without
additional terms.

\subsection{The problem with the negative energy states}

\noindent In the just raising quantum electrodynamics, the matter
fields, describing e.g. electrons, were of course assumed to be
bispinors to which the successful Dirac theory of spin 1/2
particles applied \cite{Dirac1928ab}. This brought to the quantum
electrodynamics theory, in addition to the known problem of an
infinite self-energy term for the electron, the puzzle which
seemed intrinsically associated with any relativistic theory (and
the Dirac theory of electrons in particular), i.e. that of the
negative energy states and the possible transitions from positive
to negative energy states. Heisenberg and Pauli simply discarded
such a problem, regarding the negative energy states as an
``inconsistency of the theory [...], which must be accepted as
long as the Dirac difficulty is unexplained'' \cite{HP1929a}. On
the other hand, also Fermi recognized that this ... %
\begin{quote}
... theory has two fundamental defects that, rather than of
electrodynamic origin, may be considered as coming from the
incomplete knowledge of the electron structure. They are the
possibility that the Dirac electron has to jump to negative energy
levels and the fact that the self-energy takes an infinite value
by assuming an exactly pointlike electron \cite{FNM64}.
\end{quote}
For our purposes, we definitely do not discuss further about the
last point.
\\
Instead, regarding the problem with the negative energy states, it
was soon well clear to Dirac himself that from his equation it
follows that a negative energy electron moves in an external field
as if it had positive charge and energy. This original observation
led Hermann Weyl in 1929 to ``expect that, among the two pairs of
components of the Dirac [wavefunction]  one pair corresponds to
the electron, while the other to the proton'' \cite{Weyl1929}.
This association was further considered (for some time) by Dirac
in the framework of his ``hole theory'', with the identification
of ``the holes in the distribution of negative energy electrons''
with the protons \cite{Dirac1929}, although novel difficulties
arose with such interpretation. Here we only quote from the
Dirac paper appeared at the very beginning of 1930: %
\begin{quote}
Can the present theory account for the great dissymmetry between
electrons and protons, which manifest itself through their
different masses and the power of protons to combine to form
heavier atomic nuclei? It is evident that the theory gives, to a
large extent, symmetry between electrons and protons [...] The
symmetry is not, however, mathematically perfect when one takes
interaction between the electrons into account. [...] The
consequences of this dissymmetry are not very easy to calculate on
relativistic lines, but we may hope it will lead eventually to an
explanation of the different masses of proton and electron
\cite{Dirac1929}.
\end{quote}
As well known, such difficulty in the interpretation of the
theory, and some others related to it (i.e., the too high rate of
annihilation of electrons and protons), were later overcome by
Dirac himself, after the Weyl proof that holes necessarily
represented particles with the same mass as an electron
\cite{Weyl1931}. ``A hole, if there were one, would be a new kind
of particle, unknown to experimental physics, having the same mass
and opposite charge to an electron. We may call such a particle an
anti-electron'' \cite{Dirac1931}. This novel interpretation,
though highly controversial, was eventually confirmed by the
cosmic-rays experiments by Carl D. Anderson \cite{Anderson} in
1932. However, the fact that the observed positive electrons were
indeed Dirac antiparticles was not clear for some time, but was
fully recognized only after the appearance of the experimental
results on cosmic-rays showers by Patrick M. Blackett and Giuseppe
P.S. Occhialini \cite{Blackett}, followed by a discussion of their
results within the framework of the Dirac theory. Notwithstanding
this, especially Pauli was very critical with the Dirac theory of
negative energy states. ``I do not believe in your perception of
`holes', even if the existence of the `antielectron' is proved'',
wrote Pauli to Dirac on May 1, 1933 \cite{Pauli1985}. And, more
specifically, in a letter to Heisenberg of July 24, 1933, Pauli
pointed out that the ...
\begin{quote}
... play with infinite concepts in the present framework is
unacceptable and will eventually lead to contradictions, In
particular, I don't see how the Coulomb interaction energy can be
eliminated unequivocally and free from arbitrariness with the
infinitely many occupied states \cite{Pauli1985}.
\end{quote}
An attempt was made by Heisenberg (in a letter to Pauli of July
17, 1933 \cite{Pauli1985}) to fold hole theory into the
Heisenberg-Pauli formalism of quantum electrodynamics, in order to
try to take into some account Pauli criticism. But the most
important step in this direction was carried out by Vladimir Fock,
Wendell Furry and Julius R. Oppenheimer in 1934. Inspired by the
Dirac idea to consider an annihilation as a lack of one electron
with negative energy, they defined the creation and annihilation
operators $b^\ast,b$ of the antielectrons as the corresponding
operators of annihilation and creation $a,a^\ast$ of the electrons
with positive energy. The Dirac field $\psi$ may then be expanded
as two sums over normal modes with positive and negative energies,
in such a way that the energy of the system, %
\be %
H = \sum_k^{E_k>0} \! E_k \, a^\ast_k a_k + \sum_k^{E_k<0} \!
|E_k| \, b^\ast_k b_k + E_0, \label{e3}
\ee %
when the (infinite) vacuum energy $E_0$ is subtracted, is the sum
of two terms that are positive definite.

\subsection{Quantization of the Klein-Gordon equation}

\noindent The problem of the negative energy states was solved,
for a charge particle with spin 0, in 1934 by Pauli and Victor F.
Weisskopf \cite{PW}. At the time their work was viewed just as a
theoretical exercise (the authors themselves considered their
theory to be truly a ``curiosity'', thus choosing to publish it in
a not first-line journal \cite{Miller}), since no ``elementary''
spin 0 particle\footnote{Of course, spin 0 particles were known
since 1920s, as for example the hydrogen atom in its ground state
or the helium nucleus. However neither of these particles were
``elementary'', so that no need was felt to describe them by means
of a relativistic wave-equation.} was then known. Nevertheless, as
we will see below, it proved to be a very important step in the
development of the quantum theory of electrodynamics.

The Pauli-Weisskopf paper ``consistently applies the
Heisenberg-Pauli formalism of the quantization of wave fields to
the scalar relativistic wave-equation for matter fields when the
particles have Bose-Einstein statistics'' \cite{PW}. The scalar
relativistic wave-equation was that earlier discovered by
Schr\"odinger and, independently, by Klein and Walter Gordon (for
a discussion see e.g. \cite{Miller}): %
\be %
\left[ \frac{\hbar^2}{c^2} \frac{\partial^2 ~}{\partial t^2} -
\hbar^2 \nabla^2 + m^2 c^2 \right] \psi = 0. \label{e4}
\ee %
The original rejection by Schr\"odinger (and others) of this
equation was due to its incorrect predictions, with respect to the
experimental observations, about the fine structure of the
spectrum of the hydrogen atom. This was correctly ascribed to the
fact that such an equation did not take into account the electron
spin. However the later appearance of the Dirac theory for the
spinning electron, which gave correct experimental predictions,
shifted the attention to the problem of non positive definite
probabilities, found in the Schr\"odinger-Klein-Gordon theory but
not in the Dirac theory. This brought to the generally-accepted
assumption about  the non-existence of spin 0 (elementary)
particles, which was reinforced by the interpretation underlying
the hole theory. In fact, the exclusion principle that played a
key role in that interpretation could not be applied to scalar
particles, so that no mechanism existed that prevented the
particles to pass from positive energy states to negative ones. It
is a merit of the Pauli-Weisskopf theory to have shown that non
intrinsically wrong arguments lie in the
Schr\"odinger-Klein-Gordon theory that justifies the derivation of
the Dirac theory; they just apply to particles with different
spin, and the interpretation of antiparticles in terms of holes is
no longer necessary.

As announced in the abstract of their paper, Pauli and Weisskopf
adopted the canonical formalism introduced earlier by Heisenberg
and Pauli, and applied it first of all to the quantization of the
wave field in the force-free case. By starting from the Lagrangian
density for the complex scalar field $\psi$ and its conjugate
$\psi^\ast$, they deduced the momentum $\Pi$ and $\Pi^\ast$, which
are canonically conjugate to $\psi$ and $\psi^\ast$, and imposed
the commutation relations (\ref{e2}), appropriate for the
Bose-Einstein statistics. The expressions for the Hamiltonian $H$,
charge density $\rho$ and current density $\sbvec$ operators
were obtained and, by decomposing the fields into hermitian parts, %
\be %
\psi = \frac{1}{\sqrt{2}} \left( u_1 + i \, u_2 \right), \qquad
\qquad \Pi = \frac{1}{\sqrt{2}} \left( p_1 - i \, p_2 \right)
\label{e5}
\ee %
(and similarly for $\psi^\ast$ and $\Pi^\ast$), they showed that
the charge situated within an arbitrary finite region $V$ %
\be %
\int_V \rho \, \drm V = \int_V \left( p_1 u_2 - p_2 u_1 \right) \,
\drm V , \label{e6}
\ee %
measured in units of electron charge $e$, has the eigenvalues
$0,\pm 1, \pm 2 , \ldots , \pm N$. Then Pauli and Weisskopf
considered a complete system of orthonormalized eigenfunctions
(plane waves) in the momentum space, and expanded the field
functions in terms of these: %
\be %
\ba{rcl} \displaystyle \psi &=& \displaystyle \frac{1}{V^{1/2}}
\sum_k q_k \, \erm^{i \kbvec \cdot \xbvec} , \\ & & \\
\displaystyle \Pi &=& \displaystyle \frac{1}{V^{1/2}} \sum_k p_k
\, \erm^{-i \kbvec \cdot \xbvec}  \ea \label{e7}
\ee %
(and similarly for $\psi^\ast$ and $\Pi^\ast$). The
(non-hermitian) q-numbers $q_k, p_k, q^\ast_k, p^\ast_k$, or the
annihilation and creation operators $a_k, b_k, a^\ast_k, b^\ast_k$
derived by them, satisfy canonical commutation relations, and
these were used to write the total energy $H$ and charge $Q$ (and
momentum and current) as follows:
\be %
H = \sum_k \left( p^\ast_k p_k + E^2_k \, q^\ast_k q_k \right) =
\sum_k E_k \left( a^\ast_k a_k + b^\ast_k b_k + 1 \right) ,
\label{e8} \ee%
\be%
Q = - i \sum_k \left( p_k q_k - p^\ast_k q^\ast_k \right) = \sum_k
\left( a^\ast_k a_k - b^\ast_k b_k \right). \label{e9}
\ee %
Here $E_k = + c \sqrt{\hbar^2 k^2 + m^2 c^2}$ is always positive,
and the same applies to the total energy $H$ regardless of the
zero-point energy (the term corresponding to $+1$ in Eq.
(\ref{e8})), this having a ``decisive physical meaning''.
Furthermore, it was shown that the operators $N^+_k = a^\ast_k
a_k$ and $N^-_k = b^\ast_k b_k$, appearing above, have
non-negative integer eigenvalues: ``$N^+_k$ signifies the number
of particles with charge number $+1$ and momentum $\hbar \kbvec$,
and $N^-_k$ signifies the number of particles with charge number
$-1$ and momentum $- \hbar \kbvec$''. The existence of two
different kinds of operators $a,b$ playing the same role in the
Hamiltonian, made evident the fact that the Pauli-Weisskopf theory
is a quantum theory for two scalar particles with the same mass,
these being interpreted as the particle and the corresponding
antiparticle.

The next step was to consider the interaction of the scalar
particle with an electromagnetic field, described by the
potentials $A_\mu$. This was achieved simply by the replacement
\be%
\frac{\partial \psi}{\partial x^\mu} \ \longrightarrow \
\frac{\partial \psi}{\partial x^\mu} - \frac{i \, e}{\hbar \, c}
\, A_\mu \, \psi \label{e10}
\ee%
(and similarly for $\psi^\ast$) in the original Lagrangian
function, and adding to this the appropriate kinetic term for the
electromagnetic field. The canonical formalism was then applied to
the present case, and the important result was obtained that ``the
eigenvalues of the [novel operator for the] charge density remain
the same as in the force-free case, even when external potentials
are present'', so that the previous basic interpretation of the
theory remained unchanged. The complete Hamiltonian function was
divided into three terms, pertaining to the matter fields, the
radiation field and the interaction between them, respectively.
This was later exploited in order to calculate the relevant
effects, such as pair annihilation and creation by light quanta
(and the polarization of the vacuum). ``The frequency of these
processes proves to be of the same order of magnitude as the
frequency for particles of the same charge and mass, which follows
from the Dirac hole theory''.

\

Our account of the known story of quantum electrodynamics stops
here. This is certainly not due to the lack of further
cornerstones, which came out especially after the Second World
War, but because the further developments simply focused on (very
important) formal improvements of the theory and their physical
interpretation (recall also that, in our discussion, we have
completely disregarded the problems with the infinities). With the
Pauli-Weisskopf theory a key point has been reached, with the
proof of the most important fact that antiparticles were not just
a peculiarity of Dirac theory for spin 1/2 fermions, but also a
feature of a quantum field theory for charged bosons. As well
important was that the Pauli-Weisskopf paper corrected the
generally-accepted view that the inclusion of the principles of
the theory of special relativity into quantum mechanics required
necessarily a spin 1/2, while it showed that scalar quantum
electrodynamics was able to predict physical processes similar to
spin 1/2 quantum electrodynamics.

\section{An unknown contribution by Ettore Majorana}

\noindent According to Edoardo Amaldi,

\begin{quote}
during the winter of 1928-29 Fermi started studying the quantum
theory of radiation. [...] His formulation formed the subject of a
course given in April 1929 at the Institut Poincar\'e in Paris and
in a more complete form  at the Summer School of Theoretical
Physics at Ann Arbor, Michigan during the summer of 1930. [...]
While doing this work Fermi taught his result to several of his
pupils and friends including Amaldi, Majorana, Racah, Rasetti and
Segr\`e. Every day when work was over he gathered the various
people mentioned above around his table and started to elaborate
before them, first the basic formulation of quantum
electrodynamics and then, one after the other, a long series of
applications of the general principles to particular physical
problems \cite{AmaldiFNM50}.
\end{quote}
The study of such a subject particularly impressed, among the
others, Ettore Majorana, this being widely recorded in his
personal study notes, the ``Volumetti'' \cite{Volumetti}, and
especially in his research notes, the ``Quaderni''
\cite{Quaderni}. These treasures of theoretical physics, conserved
at the Domus Galilaeana in Pisa (Italy), have been made available
to a general audience only recently with the mentioned
publications, and are bright confirmations of his high talent as
theoretical physicist, as acknowledged in several occasions by
Fermi, Heisenberg and many others (see, for example, Ref.
\cite{Recami}).

Here we concentrate only on some interesting results obtained by
Majorana several years {\it before} the appearance of the
Pauli-Weisskopf paper. The original material discussed below is
contained in the Quaderno No. 2, which is not explicitly dated.
However an accurate inspection of the arguments present in it and
the comparison with the existing literature (as, for example, the
Franco Rasetti's important work on the Raman spectra of molecules
\cite{Rasetti}) and with the Volumetto III (which is dated 1929)
allows to conclude that such work was carried out at the end of
1929 or early in 1930. This fits well with the independent piece
of information by Amaldi about the Fermi private course on quantum
electrodynamics (and with another news reported below).

Before entering into the details of the theory of Majorana, we
disclose the results achieved, by reporting what seems the only
testimony on these results (Majorana never published them), as
recalled by Gian Carlo Wick.\footnote{Wick talked several times
about the following episode (see, for example, \cite{Recami},
\cite{WickAcca}); in the following we report an excerpt from Ref.
\cite{WickAcca} and an handwritten note by Wick conserved at the
Wick Archive of the Scuola Normale Superiore in Pisa. This last
paper contains many interesting details, lacking in others
testimonies.}
\begin{quote}
[...] I was admitted as a bystander to an international meeting on
Nuclear Physics, which was organized ny the Royal Academy in Rome
in the fall of 1931. [...]

I was asked by Heitler to act as a sort of interpreter between him
and Majorana. He spoke hardly any Italian, and Majorana's German
was a bit weak. So during lunches, Heitler expressed a curiosity
in what Majorana was doing; Fermi must have told him how bright he
was. So Majorana began telling, in that detached and somewhat
ironical tone, which was typical of him, especially when
discussing his own work, that he was developing a relativistic
theory for charged particles. It was not true, he said, that the
Schr\"odinger equation for a relativistic particle had to have the
form indicated by Dirac. It was clear by n ow, that in a
relativistic theory one had to start from a field theory and for
this purpose the Klein-Gordon wave-equation was just as legitimate
as Dirac's. If one took that, and quantized it, one got a theory
with consequences quite similar to Dirac positron theory, with
positive and negative charges, the possibility of pair creation
etc. You can see what I am driving at: Majorana had the
Pauli-Weisskopf scheme all worked out already at that time. Please
do not think I am disputing the merit of these authors. Majorana
never published this work, he did not seem to take it very
seriously, and I don't think Pauli and Weisskopf ever even heard
of it. Heitler probably forgot all about it, and so did I, until I
saw the paper by Pauli and Weisskopf.

I must confess that I knew enough to understand the gist of what
he was saying, but not enough to appreciate how novel and original
it was. Heitler probably did, because his comment, as I recall it,
was: `I hope you will publish this'. [...] [But] in the case I
have described publication never occurred .
\end{quote}

\subsection{Majorana theory -- first part}

\noindent In his notebooks, Majorana started to consider scalar
quantum electrodynamics with two preliminary studies on the
quantization of the free (without interaction with the
electromagnetic field) Klein-Gordon equation and on the
quantization of the free electromagnetic field. Here we do not go
in any detail about these preliminary studies (the complete
treatment of the problem may be found in Sect.s 2.7, 2.8 and 7.6
of Ref. \cite{Quaderni}), but immediately report on the first
theoretical framework considered by Majorana.

The starting point was the following variational principle for a
complex scalar field $\psi$ and its complex conjugate $\ov{\psi}$
interacting with the electromagnetic potentials $\varphi,
\Abvec$: %
\be %
\ba{l} \dps \!\!\!\!\!\! \delta \int \left\{ \frac{h^2}{8 \pi^2 m}
\left[ \frac{1}{c^2} \left( \frac{\partial}{\partial t} + \frac{2
\pi i}{h} e \, \varphi \right) \ov{\psi} \ \left( \frac{\partial
}{\partial t} - \frac{2 \pi i}{h} e \, \varphi \right) \psi \right. \right.  \\
\\
\dps ~~~~~~~~~~~~~ \left. \left. - \left( \gr - \frac{2 \pi i}{hc}
e \, \Abvec \right) \ov{\psi} \cdot \left( \gr + \frac{2 \pi
i}{hc} e \, \Abvec \right) \psi \right] - \frac{1}{2} mc^2
\ov{\psi} \psi \right\} \drm \tau =0 . \ea \label{ee1}
\ee %
Here $-e,m$ are the electric charge and mass of the field $\psi$,
respectively, while $\drm \tau = \drm t \, \drm V$ is the 4-volume
element. By taking variations with respect to $\psi, \ov{\psi}$
one obtains the wave-equations:%
\bea %
& & \!\!\!\!\!\!\!\!\!\! \left[ \frac{h^2}{8 \pi^2 mc^2} \left(
\frac{\partial}{\partial t} - \frac{2 \pi i}{h} e \, \varphi
\right)^2 - \frac{h^2}{8 \pi^2 m} \left( \gr + \frac{2 \pi i }{hc}
e \, \Abvec \right)^2 + \frac{1}{2} mc^2 \right] \psi = 0 ,
\label{ee2a} \\
& & \!\!\!\!\!\!\!\!\!\! \left[ \frac{h^2}{8 \pi^2 mc^2} \left(
\frac{\partial}{\partial t} + \frac{2 \pi i}{h} e \, \varphi
\right)^2 - \frac{h^2}{8 \pi^2 m} \left( \gr - \frac{2 \pi i }{hc}
e \, \Abvec \right)^2 + \frac{1}{2} mc^2 \right] \ov{\psi} =0 .
\label{ee2b}
\eea%
The field $P$ canonically conjugate to $\psi$ is%
\be %
P = \frac{h^2}{8 \pi^2 m c^2} \left( \frac{\partial}{\partial t}+
\frac{2 \pi i}{h} e \, \varphi \right) \ov{\psi} , \label{ee3}
\ee %
and similarly for $\ov{P}$. The Hamiltonian of the system
considered was obtained to be:
\bea %
H &=& \int \left\{ \frac{h^2}{8 \pi^2 m c^2} \frac{\partial
\ov{\psi}}{\partial t} \frac{\partial \psi}{\partial t} +
\frac{1}{2} m c^2 \ov{\psi} \psi + \frac{h^2}{8 \pi^2 m} \gr
\ov{\psi} \cdot \gr \psi \right. \nonumber \\
& & \left. + \frac{h e}{4 \pi i m c} \left[ \ov{\psi} \Abvec \cdot
\gr \psi - \psi \Abvec \cdot \gr \ov{\psi} \right] + \frac{e^2}{2
m c^2} \left( A^2 - \varphi^2 \right) \ov{\psi} \psi \right\} \drm
V
\nonumber \\
&=& \int \left\{ \frac{8 \pi^2 m c^2}{h^2} \ov{P} P + \frac{2 \pi
i}{h} \, e \varphi \left( \psi P - \ov{\psi} \, \ov{P} \right) +
\frac{1}{2} m c^2 \ov{\psi} \psi + \frac{h^2}{8 \pi^2 m} \gr
\ov{\psi} \cdot \gr \psi \right. \nonumber \\
& & \left. + \frac{h e}{4 \pi i m c} \left[ \ov{\psi} \Abvec \cdot
\gr \psi - \psi \Abvec \cdot \gr \ov{\psi} \right] + \frac{e^2}{2
m c^2} |A|^2 \ov{\psi} \psi \right\} \drm V , \label{ee4}
\eea %
while the total number $N$ of particles is given by%
\be %
N = \int - \frac{2 \pi i}{h} \left( \psi P - \ov{\psi} \, \ov{P}
\right) \drm V . \label{ee5}
\ee %
Quantization was achieved by imposing the following conditions
(with appropriate interpretation of the symbols; we prefer to
adhere to the original notation, not intended for publication, as
long as the meaning is sufficiently clear): %
\be %
\left[ \psi , P \right] = 1, \qquad \qquad \left[ \ov{\psi} ,
\ov{P} \right] = 1 \label{ee6}
\ee %
(while the other commutators vanish). At this point Majorana
introduced the transformations: %
\be%
\psi = \frac{\psi_0 - i \psi_1}{\sqrt{2}}, \qquad \qquad \psi =
\frac{P_0 + i P_1}{\sqrt{2}}, \label{ee7}
\ee %
and similarly for $\ov{\psi}, \ov{P}$, so that the quantization
conditions change as follows: %
\be %
\ba{c} \dps \left[ \psi_0 , P_0 \right] = 1, \qquad \qquad \left[
\psi_1 , P_1 \right] = 1, \\ \\ \dps \left[ \psi_0 , P_1 \right] =
\left[ \psi_0 , \psi_1 \right] = \left[ P_0 , P_1 \right] = \left[
\psi_1 , P_0 \right] = 0 . \ea \label{ee8}
\ee %
The novel fields are then expanded in plane waves:%
\be %
\ba{rcr} \dps \psi_0 = \sum_r q_0^r u^r, & \qquad & \dps \psi_1 =
\sum_r q_1^r u^r, \\ & & \\
\dps P_0 = \sum_r p_0^r u_r, & \qquad & \dps P_1 = \sum_r p_1^r
u_r, \ea \label{ee9}
\ee %
where $q_0, q_1, p_0, p_1$ are four hermitian operators, and the
functions $u_r$ satisfy: %
\be %
\la u_r + k_r^2 u_r = 0, \qquad \qquad \int u_r^2 \drm V = 1.
\label{ee10}
\ee %
From now on (for this first part of the Majorana theory), the
discussion was specialized to consider only the case of scalar
fields without interaction with the electromagnetic field. The
Hamiltonian takes then the simpler form (here and below, for
simplicity, we suppress the indices $r$ if no confusion may be
generated): %
\bea%
\!\!\!\!\!\!\!\!\!\!\!\!\!\!\! H_0 &=& \sum \left\{ \frac{4 \pi^2
m c^2}{h^2} p_0^2 + \left( \frac{h^2 k^2}{16 \pi^2 m} +
\frac{1}{4} m c^2 \right) q_0^2
\right. \nonumber \\
& & \left. ~~~~~~~~~ + \frac{4 \pi^2 m c^2}{h^2} p_1^2 + \left(
\frac{h^2 k^2}{16 \pi^2 m} + \frac{1}{4} m c^2 \right) q_1^2
\right\} . \label{ee11}
\eea%
By introducing the quantity $\nu$ defined by %
\be %
\nu^2 = \frac{m^2 c^4}{h^2} + \frac{c^2 k^2}{4 \pi^2} ,
\label{ee12}
\ee %
the total energy is then: %
\be %
E = \sum_r N_r c \sqrt{m^2 c^2 + p_r^2} = \sum N_r h \nu_r = \sum
N_r E_r , \label{ee13}
\ee %
where the total number of particles $N$, and the total charge $Z$
could be written as: %
\be %
N = \sum_r N_r, \qquad \qquad Z = \sum_r Z_r . \label{ee14}
\ee %
Here the operator $N_r$ has eigenvalues $0,1,2,\dots$ (see below)
while $Z_r$ takes the values $N_r, N_r-2, N_r - 4, \dots , - N_r$;
from $|Z_r|\leq N_r$, Majorana then pointed out that in his theory
$|Z|\leq N$.

It is worthwhile to stress the fact that Majorana thus succeeded
to decompose the charged scalar field into a sum of normal modes,
each having a relativistic energy $c \sqrt{m^2 c^2 + p_r^2}$.

A short insert is introduced at this stage, where the
electrostatic interaction through the potentials $\Abvec = 0$,
$\varphi \neq 0$ is dealt with. The scalar potential is expanded
as: %
\be %
\varphi = \sum_r \varphi_r u_r, \qquad \qquad \varphi_r = \int \!
\varphi \, u_r^2 \, \drm V , \label{ee15}
\ee %
and, by defining %
\be %
\varphi_{rs} = \int \! \varphi \, u_r u_s \, \drm V \label{ee16}
\ee %
the complete Hamiltonian is given by: %
\be %
H = H_0 - \frac{2 \pi}{h} \sum_{r,s} \varphi_{rs} \left( q_0^r
p_1^s - q_1^r p_0^s \right) . \label{ee17}
\ee %
Turning back to the main problem, the calculations were now
simplified by taking units where $h = 2 \pi$ and $\nu = 1/ 2 \pi$,
so that $h \nu = 1$. In these units, the operators $N,Z$ were
given by: %
\be %
N = \frac{1}{2} P_0^2 + \frac{1}{2} Q_0^2 + \frac{1}{2} P_1^2 +
\frac{1}{2} Q_1^2 - 1, \label{ee18}
\ee %
\be %
Z = Q_0 P_1 - Q_1 P_0 \label{ee19}
\ee %
(with obvious meaning of the changed notations). Explicit
calculations of the matrix elements of the operators
$Q_0,Q_1,P_0,P_1,$ $N,Z$ are carried out, in the chosen basis of
eigenfunctions, by starting from the following complete set of
commutation relations: %
\be%
\ba{c} \dps P_0 Q_0 - Q_0 P_0 = \frac{1}{i}, \qquad \qquad P_1 Q_1
- Q_1 P_1 = \frac{1}{i}, \\ \\ P_0 P_1 - P_1 P_0 = \dots = 0, \ea
\ee%

\

\be%
\begin{array}{lll} NP_0 - P_0 N = i Q_0, & \quad & -(ZP_0 -
P_0 Z) = -i P_1,
\\ & \\
NQ_0 - Q_0 N = -i P_0, & \quad & -(ZQ_0 - Q_0 Z) = -i Q_1,
\\ & \\
NP_1 - P_1 N = i Q_1, & \quad & -(ZP_1 - P_1 Z) = i P_0,
\\ & \\
NQ_1 - Q_1 N = -i P_1, & \quad & -(ZQ_1 - Q_1 Z) = i Q_0 .
\end{array}
\ee %
Here we report only the explicit expressions for the total number
and charge:
\be%
\small N = \left| \begin{array}{ccccccc}
0 & 0 & 0 & 0 & 0 & 0 & \ldots \\
 & & & & & & \\
0 & 1 & 0 & 0 & 0 & 0 & \ldots \\
 & & & & & & \\
0 & 0 & 1 & 0 & 0 & 0 & \ldots \\
 & & & & & & \\
0 & 0 & 0 & 2 & 0 & 0 & \ldots \\
 & & & & & & \\
0 & 0 & 0 & 0 & 2 & 0 & \ldots \\
 & & & & & & \\
0 & 0 & 0 & 0 & 0 & 2 & \ldots \\
% & & & & & & \\
\ldots & \ldots & \ldots & \ldots &\ldots & \ldots & \ldots
\end{array} \right|, \label{ee20}
\ee %
\be %
\small Z = \left| \begin{array}{ccccccc}
0 & 0 & 0 & 0 & 0 & 0 & \ldots \\
 & & & & & & \\
0 & 1 & 0 & 0 & 0 & 0 & \ldots \\
 & & & & & & \\
0 & 0 & -1 & 0 & 0 & 0 & \ldots \\
 & & & & & & \\
0 & 0 & 0 & 2 & 0 & 0 & \ldots \\
 & & & & & & \\
0 & 0 & 0 & 0 & 0 & 0 & \ldots \\
 & & & & & & \\
0 & 0 & 0 & 0 & 0 & -2 & \ldots \\
% & & & & & & \\
\ldots & \ldots & \ldots & \ldots &\ldots & \ldots &
 \ldots \end{array} \right| .  \label{ee21}
\ee %
Furthermore, the particular action of $P_0$ on the states with
given $N,Z$ is considered, as well as the possible transitions
between states with fixed $N,Z$.

The last important thing present in the first part of the Majorana
theory is the transformation from the basis where the states of
the system were labelled according to $N,Z$ to the basis
characterized by the operators: %
\be %
\frac{N+Z}{2} = L, \qquad \qquad \frac{N-Z}{2} = M \label{ee22}
\ee %
($N=L+M$, $Z=L-M$). The eigenvalues of both these operators were
$0,1,2,\dots$ and, as explicitly reported by Majorana, $M$
numbered the particles with negative charge while $L$ those with
positive charge (that is, what Dirac in 1931 named
``anti-electrons''). The rewriting of the operators $Q_0, P_0,
Q_1, P_1$ defined in the basis $N,Z$ in terms of $Q_L,P_L,Q_M,P_M$
defined in the basis $L,M$ then followed. Through these operators,
$L$ and $M$ were written as: %
\be %
L = \frac{1}{2} P_L^2 + \frac{1}{2} Q_L^2 - \frac{1}{2}, \qquad
\qquad M = \frac{1}{2} P_M^2 + \frac{1}{2} Q_M^2 - \frac{1}{2} .
\label{ee23}
\ee %
The discussion of this first part ended with the pointing out of
the particular case corresponding, in Majorana's own words, to the
``classical theory'', that is without what we name
``anti-particles''. This was achieved, for example, by assuming
$P_M = \psi_M = 0$.

\subsection{Majorana theory -- second part}

\noindent In the original manuscript, what just reported appears
to have been considered a ``preliminary'' step, in Majorana's
view, rather than a part of a complete theory. In fact at this
point the author re-starts his calculation from a variational
principle similar to that in Eq. (\ref{ee1}), with the addition of
the
proper kinetic terms for the electromagnetic field:%
\be %
\ba{l} \dps \!\!\!\!\!\! \delta \int \left\{ \frac{h^2}{8 \pi^2 m}
\left[ \frac{1}{c^2} \left( \frac{\partial}{\partial t} + \frac{2
\pi i}{h} e \, \varphi \right) \ov{\psi} \ \left( \frac{\partial
}{\partial t} - \frac{2 \pi i}{h} e \, \varphi \right) \psi \right. \right.
\\ \\ \dps ~~~~~~~~~~~~~ \left. \left. - \left( \gr - \frac{2 \pi i}{hc}
e \, \Abvec \right) \ov{\psi} \cdot \left( \gr + \frac{2 \pi
i}{hc} e \, \Abvec \right) \psi \right] - \frac{1}{2} mc^2
\ov{\psi} \psi \right.
\\ \\ \dps ~~~~~~~~~~~~~ \left.
+ \frac{1}{8 \pi} \left( \left| \frac{1}{c} \frac{\partial
\Abvec}{\partial t} + \gr \varphi \right|^2 - | \ro \Abvec|^2
\right) \right\} \drm \tau =0 . \ea \label{ee24}
\ee %
The variations taken with respect to $\psi, \ov{\psi}$ give,
obviously, equations identical to those in Eqs.
(\ref{ee2a},\ref{ee2b}), but now they were written in the
following ``mixed'' form, where all
the four fields $\psi, \ov{\psi}, P, \ov{P}$ appear: %
\be \ba{l} \dps \ba{l} \dps \left( \frac{\partial }{\partial t} -
\frac{2 \pi i}{h} e \, \varphi \right) \ov{P} = - \frac{1}{2} mc^2
\psi  + \frac{h^2}{8 \pi^2 m} \left| \gr + \frac{2 \pi i }{hc} e
\, \Abvec \right|^2 \psi , \ea
 \\ \\ \dps \ba{l} \dps
\left( \frac{\partial }{\partial t} + \frac{2 \pi i}{h} e \,
\varphi \right) P = - \frac{1}{2} mc^2 \ov{\psi} + \frac{h^2}{8
\pi^2 m} \left| \gr - \frac{2 \pi i}{hc} e \, \Abvec \right|^2
\ov{\psi} , \ea
\\ \\ \dps \ba{l}
\dps \left( \frac{\partial}{\partial t} + \frac{2 \pi i}{h} e \,
\varphi \right) \ov{\psi} = \frac{8 \pi^2 mc^2}{h^2} P , \ea
 \\ \\ \dps \ba{l}
\dps \left( \frac{\partial}{\partial t} - \frac{2 \pi i}{h} e \,
\varphi \right) \psi = \frac{8 \pi^2 mc^2}{h^2} \ov{P} . \ea
 \ea \label{ee25} \ee %
With the introduction of the quantities: %
\be \ba{rcl} \rho &=& \dps \frac{he}{4 \pi i mc^2} \left[
\ov{\psi} \left( \frac{\partial}{\partial t} - \frac{2 \pi i}{hc}
e \, \varphi \right) \psi - \psi \left( \frac{\partial}{\partial
t} + \frac{2 \pi i}{h c} e \, \varphi \right) \ov{\psi} \right] \\
\\ &=& \dps \frac{2 \pi i}{h} e \left( \psi P - \ov{\psi} \ov{P}
\right) , \\ \\
\cibvec &=& \dps - \frac{he}{4 \pi i mc} \left[ \ov{\psi} \left(
\gr + \frac{2 \pi i}{hc} e \, \Abvec \right) \psi - \psi \left(
\gr - \frac{2 \pi i}{hc} e \, \Abvec \right) \ov{\psi} \right]
\\ \\ &=& \dps - \frac{he}{4 \pi i mc} \left( \ov{\psi} \gr \psi - \psi \gr
\ov{\psi} \right) - \frac{e^2}{m c^2} \, \ov{\psi} \psi \Abvec .
\ea
\label{ee26} \ee %
The dynamical equations obtained by Eq. (\ref{ee24}) with
variations with respect to $\varphi$ and $\Abvec$ are.
respectively: %
\be \ba{l}
 \dps \frac{1}{4 \pi} \, \di \Ebvec - \rho = 0, \\ \\
 \dps \frac{1}{4 \pi c} \, \frac{\partial \Ebvec}{\partial t} -
 \frac{1}{4 \pi} \, \ro \Hbvec + \cibvec = 0,
\ea \label{ee27}\ee%
that is the usual Maxwell equations with sources ($\Ebvec, \Hbvec$
are the electric and magnetic fields, respectively).

Some calculations were, at this stage, present in the manuscript
aimed at realizing the (now) well-known difficulty, in a canonical
theory, with the search for a field canonically conjugates to
$\varphi$ (for $\Abvec$ it is just $- \Ebvec/4 \pi c$). Such
calculations end with the ``obvious'' choice $\varphi = 0$. Thus,
summing up, the degrees of freedom of the system considered are
described, in the theory elaborated by Majorana, by the
(independent) fields: %
\be
 \psi, \quad \ov{\psi}, \quad \Abvec ,
\label{ee28} \ee %
and their canonical conjugates: %
\be P = \frac{h^2}{8 \pi^2 m c^2} \, \frac{\partial \psi}{\partial
t}, \quad \ov{P} = \frac{h^2}{8 \pi^2 m c^2} \, \frac{\partial
\ov{\psi}}{\partial t}, \quad - \frac{\Ebvec}{4 \pi c} =
\frac{1}{4 \pi c^2} \, \frac{\partial \Abvec}{\partial t},
\label{ee29} \ee %
while the Hamiltonian of the system is: %
\bea \!\!\!\!\! H & \!\! = \!\! & \int \left[ \frac{8 \pi^2
mc^2}{h^2} \ov{P} P + \frac{1}{2} mc^2 \, \ov{\psi} \psi +
\frac{h^2}{8 \pi^2 m} \gr
\ov{\psi} \cdot \gr \psi \right.  \nonumber \\
\!\!\!\!\! & & \!\! + \left. \frac{he}{4 \pi i mc} \Abvec \cdot
(\ov{\psi} \gr \psi - \psi \gr \ov{\psi} )  + \frac{c^2}{2 mc^2}
|\Abvec|^2 \ov{\psi} \psi + \frac{1}{8 \pi} (E^2 + H^2) \right]
\drm V . \label{ee30} \eea %
Such fields were, then, expanded in terms of three sets of plane
waves (one set of scalar and two of vector quantities) as follows.
Let us consider two sets of vector functions $\fbvec_\l,
\fbvec'_k$ labelled by the indices $\l, k$: %
\be %
\ba{c} \dps \la \fbvec_\lambda + \lambda^2 \fbvec_\lambda =0 ,
\qquad \la \fbvec'_k + k^2 \fbvec'_k = 0 , \\ \\
\dps \int \! \fbvec_\lambda \cdot \fbvec_{\lambda'} \ \drm V =
\delta_{\lambda \lambda'} , \qquad \int \! \fbvec'_k \cdot
\fbvec'_{k'} \ \drm V = \delta_{k k'} , \qquad  \int \!
\fbvec_\lambda \cdot \fbvec'_k \ \drm V = 0 . \ea
\label{ee31} \ee %
By imposing the constraint $\ro \fbvec_\l = 0$, the quantities
$\fbvec_\l$ may be written in terms of the gradient of some
functions, $\fbvec_\l = \gr u_\l / \l$, where %
\be %
\la u_\lambda + \lambda^2 u_\lambda =0 , \qquad
 \int \! u_\lambda \cdot u_{\lambda'} \ \drm V = \int \!
\fbvec_\lambda \cdot \fbvec_{\lambda'} \ \drm V = \delta_{\lambda
\lambda'} .
\label{ee321} \ee %
In terms of these functions, the fields $\psi, P$ were written as: %
\be \begin{array}{l} { \dps \psi = \frac{1}{\sqrt{2}} \sum_\lambda
\sqrt{\frac{mc}{\sqrt{m^2c^2 + \lambda^2 h^2 /4 \pi^2}}} \left[
q_\lambda + q'_\lambda + i \left(p_\lambda -p'_\lambda \right)\right] u_\lambda , } \\ ~~\\
{\dps P = \frac{h}{4 \pi\sqrt{2}} \sum_\lambda \sqrt{\frac{
\sqrt{m^2c^2 + \lambda^2 h^2 /4 \pi^2} }{mc}} \left[p_\lambda +
p'_\lambda + i \left(q_\lambda -q'_\lambda \right)\right]
u_\lambda ,}
\end{array}  \label{ee33} \ee %
and similarly for $\ov{\psi}, \ov{P},$ where the four operators
$q_\lambda, q'_\lambda, p_\lambda, p'_\lambda$ satisfy the
commutation relations: %
\be \ba{c} \dps
 p_\lambda q_\lambda - q_\lambda p_\lambda = \frac{1}{i}
\qquad p'_\lambda q'_\lambda - q'_\lambda p'_\lambda = \frac{1}{i} , \\ \\
\dps p_\lambda q'_\lambda - q'_\lambda p_\lambda = p'_\lambda
q_\lambda - q_\lambda p'_\lambda = p_\lambda p'_\lambda -
p'_\lambda p_\lambda = q_\lambda q'_\lambda - q'_\lambda q_\lambda
= 0 .
\ea \label{ee34} \ee %
For the electromagnetic field Majorana instead obtained: %
\be \begin{array}{rcl} \dps \Abvec &=& \dps \sum_k \sqrt{\frac{2
hc}{k}} \ Q_k \ \fbvec'_k + \sum_\lambda \frac{hc}{\sqrt{\pi}} \
P_\lambda \ \fbvec_\lambda , \\ & & \\
\dps - \Ebvec &=& \dps \sum_k \sqrt{2 hc k} \ P_k \ \fbvec'_k -
\sum_\lambda \sqrt{4 \pi} \ Q_\lambda \ \fbvec_\lambda ,
\end{array} \label{ee35} \ee %
with %
\be  P_k Q_k - Q_k P_k = \frac{1}{i} , \qquad P_\lambda Q_\lambda
- Q_\lambda P_\lambda = \frac{1}{i} ,
\label{ee36} \ee %
the other commutators vanishing.\footnote{Formally, even for the
electromagnetic field Majorana considered {\it four} operators
$Q_k, Q_\l, P_k, P_\l$ instead of only two (for example $Q_,
P_k$); but, as we will see below, the number of free photons was
given in terms of only $Q_, P_k$.} The total charge of the system
is given by %
\be - Z \, e = \int \rho \ \drm V = \frac{2 \pi i}{h} \ e \int
\left( \psi P - \ov{\psi} \ov{P} \right) \drm V ,
\label{ee37} \ee %
where %
\bea Z & = & \sum_\lambda \left[ \left( \frac{1}{2} p_\lambda^2 +
\frac{1}{2} q_\lambda^2 - \frac{1}{2}\right) - \left( \frac{1}{2}
p^{'2}_\lambda + \frac{1}{2} q^{'2}_\lambda - \frac{1}{2} \right)
\right] \\ & = & \sum_\lambda (N_\lambda - N'_\lambda) =
\sum_\lambda Z_\lambda , \label{ee38} \eea %
with obvious meaning of the number operators $N_\l$ and $N'_\l$.
The number of the quanta of the electromagnetic field (photons)
with frequency $\nu_k = c k / 2 \pi$ is instead given by: %
\be {N}_k = \frac{1}{2} (P_k^2 + Q_k^2) - \frac{1}{2} .
\label{ee39} \ee The Hamiltonian of the complete system was
written as %
\be {H} = {H}_M + {H}_R \label{ee40} \ee %
where %
\be H_R = \sum_k {N} h \nu_k  \ + \ \sum_\lambda \frac{1}{2}
Q_\lambda^2 \ + ~\mbox{zero point energy} \label{ee41}
\ee %
is the term describing radiation, and $H_M = H_M^0 + H_M^1$ with %
\be H^0_M = \sum_\lambda (N_\lambda + N'_\lambda) c \,
\sqrt{m^2c^2 + \lambda^2 \frac{h^2}{4 \pi^2}} \ + \ \mbox{zero
point energy}  \label{ee42} \ee %
the free (matter) particle hamiltonian term, and $H_M^1$ describes
the interaction among particles and between particles and light
quanta.

Although an explicitly written physical interpretation of the
formalism developed was not reported in the manuscript by Majorana
(compare with what briefly discussed above in Sect. 1.4), it seems
very clear that an independent (and, to some extent, different)
formulation of the Pauli-Weisskopf theory was just formulated by
the Italian physicist.

The original manuscript proceeded with some other calculations;
here we point out only the following interesting things.

A transformation for the operators $q_\l, q'_\l, p_\l, p'_\l$ was
introduced (apparently) in order to simplify the expressions in
Eqs. (\ref{ee33}): %
\be
\begin{array}{lcl}
{\dps a_\lambda = \frac{1}{\sqrt{2}} (q_\lambda + i p_\lambda)} ,
&~& {\dps b_\lambda = \frac{1}{\sqrt{2}} (q'_\lambda + ip'_\lambda)} , \\ & & \\
{\dps \ov{a}_\lambda = \frac{1}{\sqrt{2}} (q_\lambda - i
p_\lambda)} , &~& {\dps \ov{b}_\lambda = \frac{1}{\sqrt{2}}
(q'_\lambda - ip'_\lambda)} . \end{array} \label{ee43} \ee %
The interesting point here is that the relations corresponding to
Eqs. (\ref{ee34}), %
\be \ba{c} \dps \left[a_\lambda, \ov{a}_\mu\right] -
\left[\ov{b}_\lambda, b_\mu\right] = 2 \delta_{\lambda \mu} ,
\qquad \dps \left[\ov{a}_\lambda, a_\mu\right] - \left[b_\lambda,
\ov{b}_\mu\right] = - 2 \delta_{\lambda \mu} , \\ \\
\dps [a,a] = [b,b] =[a,b] = [b,a] = 0 ,
\ea \label{ee44} \ee %
were written with the explicit notation %
\be \left[ x, y \right] = x \ y \mp y \ x \ee %
where the upper/lower sign referred to Bose-Einstein/Fermi-Dirac
particles. Evidently Majorana was well conscious that the
fundamentals of his theory applied to bosons as well as to
fermions.

Other lengthy calculations were carried out in the original
manuscript aimed at the quantization of the free radiation field
in a different basis of plane waves (what we may call an helicity
basis). Explicit expressions for the matrix elements of the
operators appearing in the $\Abvec, \Ebvec$ fields, the energy
density $(\Ebvec^2 + \Hbvec^2) / 8 \pi$ and the angular momentum
density $(\Ebvec \times \Hbvec)/ 4 \pi$ were reported (we address
the interested reader to Sect.s 2.9-2.11 of Ref. \cite{Quaderni}).

\subsection{Possible application to the nuclear theory}

\noindent An interesting application of what elaborated by
Majorana on quantum electrodynamics was seemingly performed by him
about the nuclear constitution. To put this into a context, first
of all we take into account of a letter written by Majorana
himself to his friend and colleague Giovanni Gentile Jr on
December 22, 1929:
\begin{quote}
[...] I have read the article by Gamov that you pointed out to me;
it seems to me that it gives really a good idea of what are the
first stirrings of the rising theory of nuclei. This theory,
however, has apparently no probability to reach its maturity
unless we graft it onto the stump of quantum electrodynamics,
that, in its turn, still gives its most pitful cries (read e.g. an
article by Landau in the November issue of ``Physikalische'', if I
am right).

In other words, it seems to me that the problem of the aggregation
of protons and electrons in nuclei cannot have solutions, though
approximate, until the problem of the constitution of protons and
electrons themselves has been solved. This is for a very simple
reason: the size of complex nuclei, as resulting from the Gamov
theories, is of the same order of that of electrons (obviously,
this is evaluated on classical grounds. Quantum mechanics has not
brought, and {\em cannot brought by itself}, any light on this
question, since no relation between $e, h, m$ can hold due to
dimensional reasons). Although such statements are largely vague,
we can expect that something is hidden in them [...]. I am thus
lead to believe that protons and electrons interpenetrate into
nuclei in a sense different from that of wave mechanics, i.e. in a
sense that is not liable to statistical interpretation [...].
\cite{letterGG}
\end{quote}
At the time, in fact, it was commonly accepted that a nucleus
$(Z,A)$ was composed of $A$ protons and $A-Z$ electrons
\cite{pndeg}, but the observations of the rotational Raman
spectrum of gaseous nitrogen by Rasetti \cite{Rasetti},
interpreted\footnote{The nucleus of nitrogen was assumed to be
composed of 14 protons and 7 electrons, that is an odd total
number of fermions, but on the contrary to expectations it was
observed to follow the Bose-Einstein statistics, rather than the
Fermi-Dirac one.} by Walter Heitler and Gerhard Herzberg
\cite{HeitHerz}, opened a deep crisis in the current model of
nucleus, which was correctly solved only with the discovery of the
neutron in 1932. The mention above by Majorana was, then, nothing
but the Heitler-Herzberg conclusion that (according to the current
nuclear model) nuclear constituents ``loose'' their spin and the
right to determine the statistics of the nucleus.\footnote{It is
right and proper to point out that the Heitler-Herzberg analysis
of experimental data lead (correctly) to that conclusion only for
electrons but, evidently, such a dissymmetry among the nuclear
constituents (protons and electrons) sounded strange to Majorana,
as well as to many other physicists.}

Another trace of the interest of Majorana on such topics, related
to the work by Rasetti on the spectra of diatomic molecules, was
left in his Quaderni and, in particular, on his Quaderno No. 2,
where several studies on diatomic molecules appear along with the
calculations on quantum electrodynamics reported above.

Here, however, we want to focus on a possible application of the
Majorana theory in Sect. 2.3 to the nucleus, which again was given
in Quaderno No. 2, just mixed up with the developments of that
theory, apparently according to Majorana's view of ``grafting the
theory of nuclei onto the stumps of quantum electrodynamics''.

\begin{figure}
\begin{center}
\epsfxsize=10cm %
\epsffile{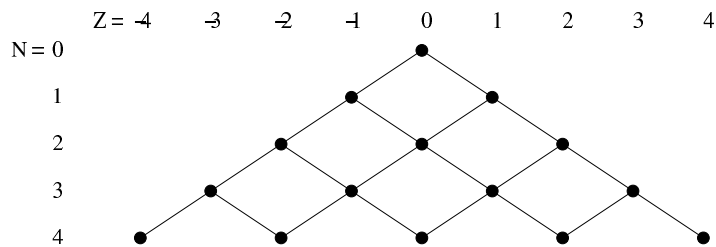} %
\end{center}
\caption{The building of nuclear states labelled by the total
number of particles $N$ and the net charge $Z$.}
\end{figure}

The two different kinds of scalar particles (endowed with positive
and negative charge, respectively), described by the complex
scalar field $\psi$ and its conjugate $P$ (together with
$\ov{\psi}, \ov{P}$ are likely assumed to be the two nuclear
constituents. The net charge was then, as usual, denoted by $Z$,
while the total number of the constituents was $N$ (instead of
$A$). Thus, in the present case, the number of protons and
electrons would be $L$ and $M$, respectively (see Eq.
(\ref{ee22})).\footnote{This notation is similar, but not
identical to that used by Fermi \cite{FNM72}, who denoted the
atomic weight by $M$.} It is not surprising such an association in
the theory elaborated, given the large difference in mass between
protons and electrons. In fact, it should be remembered that at
the time when Majorana performed these studies, the Dirac theory
of the electron still interpreted the ``holes'' of quantum
electrodynamics in terms of protons (rather than antielectrons),
despite its possible problem \cite{Weyl1929} \cite{Dirac1929}.

Assuming such an interpretation, Majorana then built up a
mass-charge $(N,Z)$ diagram for the possible nuclear states (see
Figure 1) and, more important, a scheme for the possible
transitions between nuclei of different $N,Z$ allowed in his
theory (see Figure 2). For example, in this scheme the
$\beta$-decay of a nucleus is described by a transition from $N,Z$
to $N-1,Z+1$. Majorana also calculated several matrix elements for
these transitions, as already mentioned in Sect. 2.1, but this
application was soon abandoned, probably due to the realization of
the wrong predictions of the theory both on the compositions of
nuclear states and on the transitions between them. Thus,
differently from the elaboration of a theory for scalar
electrodynamics, such an application remains just as an
interesting historical curiosity.

\begin{figure}
\begin{center}
\epsfxsize=10cm %
\epsffile{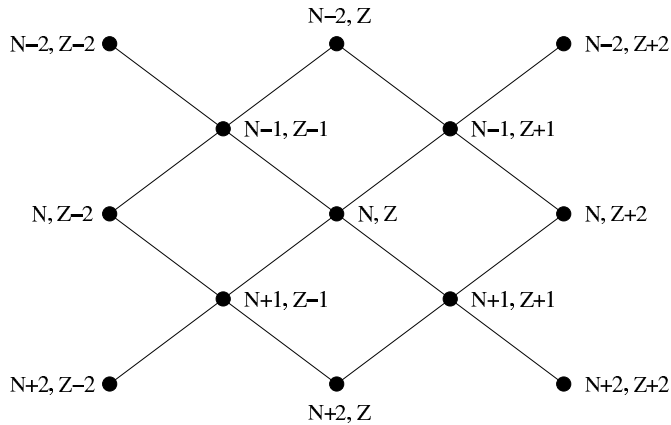} %
\end{center}
\caption{Possible transitions between nuclei with given $N,Z$.}
\end{figure}

\section{Conclusions}

\noindent The path to a complete and consistent formulation of a
quantum electrodynamics, as recalled in Sect. 1, has undoubtedly
one of its milestones in the development of the Pauli-Weisskopf
theory for a scalar field. As seen, this is not mainly related to
its possible direct applications, although the appearance of the
Pauli-Weisskopf paper is quite rightly considered as the beginning
of meson theory, but rather to the fact that it showed
unambiguously that the marriage between quantum mechanics and
special relativity did not necessarily require a spin 1/2 for the
correct interpretation of the formalism, as erroneously believed.
According to Weisskopf himself, it was a ``tremendous fun in
working out something that, at that time, was quite unexpected:
that one can get pair creation and pair annihilation without a
Dirac equation, also for particles without spin'' \cite{Wlife}.

In this respect, it is particularly relevant and interesting, from
an historical point of view, the earlier formulation of such a
theory by Majorana. This is not for a mere and quite fruitless
priority issue, but rather for the discussion and consideration of
the fundamental questions behind that theory as early as
1929-1930.

However, as should be clear from what reported in this paper, the
Majorana theory presents also interesting theoretical
peculiarities with respect to the Pauli-Weisskopf theory even for
nowadays research, such as the use of general sets of plane waves
in the expansion of the field variables or the adoption of four
($q_\l, q'_\l, p_\l, p'_\l$, for matter particles) plus four
($Q_k, P_k, Q_\l, P_\l$, for photons) instead of four plus two
operators describing the quanta of the appropriate fields.

We then expect that further progress in these studies may
certainly shed some new light on several important topics of
present-day physics.

%\acknowledgments

\subsection*{Acknowledgments}
Interesting and valuable discussions with Erasmo Recami and
Alberto De Gregorio are here kindly acknowledged.

\end{document}